\documentclass[pre,showpacs,preprint]{revtex4}
\usepackage{graphicx}
\usepackage{amssymb}
\usepackage{amsmath}
\usepackage{bm}

\begin{document}
\bibliographystyle{apsrev}

\title[]{LONG RANGE INTERACTION YIELDS A NEW KIND OF PHASE TRANSITION}

\author{Mark Ya. Azbel'}

\affiliation{School of Physics and Astronomy, Tel-Aviv University, \\
           Ramat Aviv, 69978 Tel Aviv, Israel}
\thanks{Permanent address.}
\affiliation{Max-Planck-Institute f\"{u}r Festkorperforschung - CNRS, \\
           F38042 Grenoble Cedex 9, France}

\begin{abstract}
   DNA denaturation, wetting in two dimensions, depinning of a flux
   line, and other problems map onto a phase transition with
   effective long range interaction. It yields giant non-universal
   critical indexes, arbitrarily large macroscopic correlation
   length and fluctuations at a finite distance from the critical
   temperature. In the vicinity of this region the Gibbs
   distribution is invalid, and thermodynamics must be calculated
   from the first principles. There are no fluctuations above the
   critical temperature.
\end{abstract}


\maketitle

Thermal unbinding (melting, coiling, denaturation) of a double-stranded DNA molecule is biologically important
and physically unique.  It yields a phase transition in a one-dimensional system \cite{Poland}.  The system is
extraordinary long - the total length of a single mammalian DNA is $1.8 m$, it consists of $ \sim 5$ billion nucleotide
base pairs.
Their sequence is related to genetic information, yet statistically it is close to a random one \cite{Azbel1995}.
The fraction of unbound base pairs as a function of temperature (``the DNA melting curve'')
is proportional to DNA light absorption at about $260 nm$.  DNA denaturation maps onto a variety of
other problems:  the binding transition of a polymer onto another polymer, a membrane, or an interface \cite{Fisher1984};
wetting in two dimensions \cite{Forgas};  depinning of a flux line from a columnar defect in type-II superconductors
\cite{Nelson}; localization of a copolymer at a two-fluid interface \cite{Carel}. DNA denaturation has been
extensively studied
for nearly four decades \cite{Poland, Azbel1995, Fisher1984, Forgas, Nelson, Carel, Wartell, Azbel1973,
Tang, Kafri, Fisher1966, Carlon}.
Yet, some features of this transition were overlooked.  Start with its physics
and model. DNA nucleotide base pairs (adenine - thymine AT, guanine - cytosine GC) are large (``mesoscopic'')
organic molecules.  Their unbinding releases few thousand degrees of freedom.  The corresponding entropy is
$sk_{B}$  per site \cite{Wartell} ($k_{B}$ is the Boltzmann constant, $s \sim 10$).  So, while the binding (hydrogen)
energy of DNA strands is $\sim 3000^\circ K$, DNA melts at a relatively low room temperature ($\sim 300^\circ K$), i.e.
in the vicinity of the ground state.  The Poland-Scheraga model \cite{Poland} of DNA melting introduces the fusible
AT and refractory binding
energies $E_{1} = -sk_{B} T_{1}$ and $E_{2} = -sk_{B} T_{2}$ correspondingly ($T_{1} < T_{2}$), the boundary energy
$J$ per bound segment
($J \sim 3000^\circ K$ accounts for an incomplete unbinding at the boundaries), and the loop entropy $-ck_{B} \ln L$ per
an unbound
segment ($L$ is the total number of nucleotide pairs there).  The value of the constant $c$ may vary \cite{Poland,
Wartell, Kafri, Fisher1966, Carlon}
from $1.5$ to slightly higher than $2$.  Thus, at the temperature $T$, Poland-Scheraga Hamiltonian $E_{lLx}$ of
the adjacent bound and
melted segments is related to the length $l$ and the GC concentration $x$ in the former and to the length $L$ in the
latter.  Calculated from the energy $-sk_{B} T$ per site of a completely melted DNA ($T$ is the temperature),
\begin{equation} \label{vien1}
\begin{split}
   & E_{lLx} = sk_{B} l \delta T + J + ck_{B} T \ln L - sk_{B} l (\bar{x} - x) \Delta T. \\
   & \delta T = T - \bar{T}, \quad \bar{T} = T_{1} \bar{x} + T_{2} (1- \bar{x}), \quad \Delta T = T_{2} - T_{1},
\end{split}
\end{equation}
where $\bar{x}$ is the AT concentration at an entire DNA.
Parameters $\bar{T} \sim 310K$, $\Delta T \sim 40K$ depend on the
DNA solution \cite{Wartell}. The Poland-Scheraga model
\eqref{vien1} allows for a straightforward calculation of the
homopolymer thermodynamics. The calculation yields macroscopic
fluctuations. I prove that heterogeneity just renormalizes the
interaction parameters $c$ and $J$. The renormalization depends on
temperature and heterogeneity. This implies an unusual
non-universal transition, which invalidates the Gibbs distribution
in its close, yet macroscopic vicinity. Start with the well known
case of a homopolymer \cite{Poland, Fisher1984, Wartell,
Azbel1973}. There $x = \bar{x}$, the last term in Eq.
\eqref{vien1} is missing, and $E_{lLx} = E(l,L)$ depends on $l$
and $L$ only. Then an entire Hamiltonian $H = \sum_{n} E (l^{(n)},
L^{(n)})$ describes an ideal gas of the pairs $(l^{(n)},
L^{(n)})$. It relates the free energy $f$ per site to the
normalization condition for the Gibbs probability $p_{lL}$ of
given $l$ and $L$:
\begin{equation} \label{vien2}
   p_{lL} = [ \exp {-(l + L) \phi - E (l,L) / k_{B} T}]; \quad
   \sum_{l,L=1}^{\infty} p_{lL} =1, \quad \phi = -f / k_{B} T
\end{equation}
When $\phi \ll \exp (-J / k_{B} T)$, Eqs. (\ref{vien1},\ref{vien2}) yield
\begin{equation} \label{vien3a}
   \int_{1}^{\infty} \exp (-L \phi) L^{-1-c_{1}}dL =
   (\phi + \tau) \exp (J / k_{B} \bar{T}); \quad
   \tau = s \delta T / \bar{T}; \quad c_{1} = c-1 \tag{\ref{vien3}a}
\end{equation}
Consistent with the Landau-Peierls theorem for the Hamiltonian \eqref{vien1}, when $c_{1} > 1$,  Eq.
\eqref{vien3a} yields phase transition. Then, by Eq. \eqref{vien2},  $\phi \equiv 0$ does not allow for any
excitations of a completely melted polymer. This is specific for the Hamiltonian which depends on $ \ln L$
only --- when $L = \infty$, any excitation implies an infinite energy increase.  Dependence on $ \ln L$ yields
other unusual implications also. Transition is non-universal - its critical indexes depend on $c_{1}$.
Immediately below the critical temperature \cite{Poland, Fisher1984, Wartell, Azbel1973}
$T_{c}$,
\begin{equation} \label{vien3}
\begin{split}
   & \phi = \theta \; \; \text{if} \; \; c_{1} > 1; \quad
     \phi \sim  {- \theta \ln \theta} \; \; \text{if} \; \; c_{1} = 1; \quad
     \phi \sim \theta^{1 / c_{1}} \; \; \text{if} \; \; 1 > c_{1} > 0 \\
   & \theta = (T_{c} - T) / (T_{c} - \bar{T}); \quad
     \tau_c = (T_{c} - \bar{T}) / \bar{T} = (1 / sc_{1}) \exp (-J / k_{B} T);
   \end{split}
\end{equation}
As anticipated, the critical $c_{1} = 0$, while $J / k_{B} T \sim
T/ \; \Delta T \sim s \sim 10$ implies, by Eq. \eqref{vien3}, a
very narrow width of the transition $\sim (T_{c} - \bar{T}) /
\bar{T} \sim 10^{-5}$ (i.e. $T_{c} - \bar{T} \sim 10^{-3} K$ ),
and its very close proximity to the ground state melting
temperature $\bar{T}$. Once the free energy \eqref{vien3} is
known, the Gibbs probability \eqref{vien2} allows one to calculate
any thermodynamic averages and fluctuations. The average (denoted
by a bar) relative number $\bar{ \omega } = \overline{l / L}$ of
the bounded sites, which is measured via light absorption, the
average length $\bar {L}$ of a melted segment, and their relative
mean squared fluctuations $ \Delta \omega / \bar{\omega}$, $
\Delta L / \bar{L}$ are:
\begin{subequations}
 \begin{equation} \label{vien4a}
 \begin{split}
   & \bar{\omega} \sim c_{1} \exp (J / k_{B} T) \gg 1; \quad \Delta \omega / \bar{\omega} \sim 1 \\
   & \bar{L} \sim 1 \; \; \text{if} \; \; c_{1} > 1; \quad \bar{L} \sim \phi^{c_{1}-1} \; \; \text{if} \; \; c_{1} < 1 \\
   & \Delta L / \bar{L} \sim 1 \; \; \text{if} \; \; c_{1} > 2; \quad
      \Delta L / \bar{L} \sim \phi^{0.5 c_{1} -1} \; \; \text{if} \; \; 2 > c_{1} > 1; \quad
      \Delta L / \bar{L} \sim \phi^{-0.5 c_{1}} \; \; \text{if} \; \; 1 > c_{1} > 0
   \end{split}
 \end{equation}
Thus, $\Delta \omega / \bar{\omega}$, $\Delta L / \bar{L}$ are never small, while
$\Delta L / \bar{L} \rightarrow \infty$ when $T \rightarrow T_{c}$ and $c_{1} < 2$.
A more physically meaningful fluctuation is
 \begin{equation} \label{vien4b}
   \Delta^{\ast} \omega / \bar{\omega} \equiv \overline{| \omega - \bar{\omega}|} /
       \bar{\omega} \sim \bar{\omega}^{-c_{1}} \ll 1; \quad
   \Delta^{\ast} L / \bar{L} \equiv \overline{| L - \bar{L}|} / \bar{L} \sim 1
 \end{equation}
\end{subequations}
It demonstrates, in particular, that a characteristic $|L - \bar{L}| \sim \bar{L}$ implies a characteristic
$|\ln L - \ln \bar{L}| \sim 1$ i.e., $\ll \ln \bar{L}$ when $c_{1} < 1$ and $\bar{L} \rightarrow \infty$.

Consider heterogeneous DNA. When temperature increases from $\bar{T}$ to $\bar{T} + \delta T$,
the Poland-Scheraga Hamiltonian \eqref{vien1} complements the energy increase of an ``average'' bounded segment
(the first three terms) with the energy decrease of a refractory bounded segment (the last term).
I prove that in the vicinity of the DNA melting temperature, the last term may be replaced with its
thermodynamic average for given lengths of the successive bound and unbound segments.
(Such replacement is equivalent to an unusual mean field approximation, which becomes accurate at the
phase transition and which technically reduces to a constrained summation in the partition function).
The resulting Hamiltonian describes a homopolymer with the renormalized loop entropy.
The renormalization, and thus the phase transition singularity it determines, are non-universal and depend
on the DNA parameters.

Physics of DNA melting was elucidated in
Ref.~\onlinecite{Azbel1973}. (All following statements are later
accurately verified). The segments, rich in the fusible AT, melt
first, while the richest in the refractory GC melt last. When $c >
1$ in Eq. \eqref{vien1}, bounded segments completely vanish at a
finite critical temperature $T_{c}$. There $L \rightarrow \infty$
and the effective boundary energy $(J + c k_{B} T \ln
L)\rightarrow \infty$ if $c<2$.  Then, the excitation energy also
$\rightarrow \infty$, DNA approaches its ground state,
fluctuations of $l$ and $x$ vanish, and the length of a ground
state (i.e. sufficiently refractory) bounded segment is
\cite{Azbel1973} $\propto \ln L \rightarrow \infty$, to compensate
the effective boundary energy in Eq. \eqref{vien1}. Sufficiently
close to $T_{c}$, $l$ exceeds any finite correlation length, and
the probability $w(l,x)$ of a given $x$ at such $l$ is Gaussian.
Since fluctuations of $l$ and $x$ vanish at $T_{c}$, it is $\sim$
the thermodynamic probability $l/ L$ of a bounded site. So, $l$
and $x$ yield $L(l,x)\sim l/ w(l,x)$. Thus, physics of DNA melting
suggests, and further calculation verifies, that the values of $l$
and $L \gg l$ at a given temperature determine the corresponding
value of $x$ according to $w(l,x) \sim l/ L$. The Gaussian $w$
implies $\sqrt{l} (\bar{x} - x) \propto \sqrt{\ln [1/ w(l,x)]} \gg
1$ where the factor in $w \sim l/ L$ may be disregarded with
negligible error.  Thus, $\bar{x} -x$ in Eq. \eqref{vien1} may be
replaced with its thermodynamic average according to
\begin{equation} \label{vien5a}
   l / L = (l / 2 \pi D^{2})^{1/2} \exp (-u^{2}); \quad
   u^{2} = l (\bar{x} -x)^{2} / 2 D^{2}; \quad
   D^{2} = \bar{x} (1- \bar{x}) \tag{\ref{vien5}a}
\end{equation}
In fact, large $J/ k_{B} T \sim T/ \Delta T \sim s \sim 10$ allow for Eq. \eqref{vien5a} already slightly above
$\bar{T}$. Equation \eqref {vien1}, complemented with the unusual mean field approximation \eqref{vien5a} for $x$,
yield the renormalized Hamiltonian $E^{\ast} (l,L)$, which depends on the variables $l$ and $L$ only.  In the leading
(in $l/L \ll 1$) approximation it equals
\begin{equation} \label{vien5}
   E^{\ast} (l, L) = s l k_{B} \delta T + J + c k_{B} T \ln L - s k_{B} D \Delta T \sqrt{2 l \ln L}.
\end{equation}
The last refractory term accounts for the thermodynamic average of
$x$ for given values of $l$, $L$, $\bar{x}$ and $\Delta T$ in the
Poland-Scheraga model for a heteropolymer.  The Hamiltonian
\eqref{vien5} describes a``renormalized'' homopolymer, and Eq.
\eqref{vien2}, where $E(l,L)$ is replaced with $E^{\ast} (l,L)$,
yields its exact free energy. The knowledge of the free energy
allows one to calculate all averages and their fluctuations, to
estimate the accuracy of the crucial approximation \eqref{vien5a},
to prove that its inaccuracy vanishes at the phase transition, and
thus Eq. \eqref{vien5} accurately determines transition
singularities. It also presents an exactly solvable model of a
heteropolymer and DNA. The competition in Eq. \eqref{vien5} of the
energy increase and decrease, correspondingly in the ``average''
first and last ``refractory'' terms, yields a high and relatively
narrow $ E^{\ast} (l,L)$ minimum at the ground state $l = l_{m} =
0.5 (D \Delta T/ \delta T)^{2} \ln L$ (which is indeed $\propto
\ln L$ as stated earlier). The expansion of $E^{\ast} (l,L)$ in
$(l-l_{m})$ non-universally decreases the factor $c$ in the loop
entropy by $s(D \Delta T)^{2} / 2T \delta T$,  and Eq.
\eqref{vien2}, with $E$ replaced with the expanded $E^{\ast}$,
after a straightforward calculation,  yields
\begin{equation} \label{vien6}
   \int_{1}^{\infty} (\ln L)^{1 / 2} L^{-1-\delta} \exp (-\phi L) dL = M
\end{equation}
where
\begin{equation} \label{vien7}
 \begin{split}
     \delta & = c_{1} - \gamma; \quad
        \gamma = s (D \Delta T)^{2} / 2 \bar{T} \delta T; \\
     M & = \pi^{- 1 / 2} ( 2 c_{1})^{- 3 / 2} (s D \Delta T / \bar{T})^{2} \exp (J / k_{B} \bar{T}) \gg 1.
  \end{split}
\end{equation}
Note that the left hand side of Eq. \eqref{vien6} depends on $\phi$ and $\delta$ only.  Thus, Eq. \eqref{vien6}
reduces five dimensionless parameters $(J/ T$, $\Delta T/ T$, $T/ \bar{T}$, $\bar{x}$, $c$), which determine $\phi$
in a non-renormalized case, to two parameters ($\delta$ and $M$).  When $c_{1} < 1$ and $\phi \ll \tau_{c} - \tau$,
Eq. \eqref{vien6} maps onto Eq. \eqref{vien3a}, where $c_{1}$ is renormalized into the temperature dependent
$\delta$ (which, unlike $c_{1}$, may be of any sign and which dominates the temperature dependence in the vicinity
of the phase transition).
By Eq. \eqref{vien6}, $\phi \geq 0$. Since $c_{1} > 0$ \cite{Poland, Wartell, Kafri, Fisher1966, Carlon},
$\phi = 0$ is achieved (as stated) at finite temperature $T = T_{c}$.  There
\begin{equation} \label{vien8}
  \begin{split}
    & \delta (T_{c}) = \delta_{c} = c_{1} \theta^{\ast}, \quad
       \theta^{\ast} = (2 \pi^{2})^{1 / 3} (\bar{T} / s D \Delta T)^{4 / 3} \exp (-2 J / 3 k_{B} \bar{T}) \\
    & \delta T_{c} / \bar{T} \simeq (s / 2 c_{1})(D \Delta T / \bar{T})^{2}, \quad
       \delta T_{c} \equiv T_{c} - \bar{T}
  \end{split}
\end{equation}
Note that, by Eq. \eqref{vien8},  $\delta T_{c} \sim 3^{\circ} K$.
When $T_{c} - T \ll \delta T_{c}$, then $\delta =
c_{1}(\theta^{\ast} - \theta$), where $\theta = (T_{c} - T)/
\delta T_{c}$ is the relative distance to the critical
temperature. By Eq. \eqref{vien4b}, $\delta \ll 1$ implies $\delta
\ln L \simeq \delta \ln \bar{L}$, and thus verifies the derivation
of Eq. \eqref{vien5}.  At $T_{c}$, by Eq. \eqref{vien8},
$\delta_{c} \sim 0.01$, i.e. it is very close to the critical
$c_{1} = 0 - cf$  Eq. \eqref{vien3}.  By Eq. \eqref{vien6}, $L
\propto 1/ \phi \rightarrow \infty$  when $T \rightarrow T_{c}$.
This, and $l \sim l_{m} \propto \ln L$ verify all previous
estimates.  When $\delta$, $\phi \ll 1$, asymptotics in Eq.
\eqref{vien6}, where $M \gg 1$, yield an unusual non-universal
singularity:
\begin{subequations}
\begin{align}
   & \phi \sim [ (3 \theta / 4 \pi^{1/2} \theta^{\ast}) \sqrt{\ln (\theta^{\ast} / \theta) } ]^{1 / c_{1} \theta^{\ast}};
      \quad \; \;  \text{when} \; \; \theta \ll \theta^{\ast} \label{vien9a} \\
   & \phi \sim [ (2 \theta^{\ast} / \pi \theta ) \sqrt[3]{\ln (\theta / \theta^{\ast}) } ]^{3 / 2 c_{1} \theta};
      \quad \; \; \text{when}\; \;  \theta^{\ast} \ll \theta \ll  1 \label{vien9b}
\end{align}
\end{subequations}

Consider the implications of Eqs. (\ref{vien7}--\ref{vien9b}). In
natural DNA $J / k_{B} \bar{T} \sim  \bar{T} / \Delta T \sim 10$,
$D \sim 1 / 2$, $c_{1} \sim 1$. So, in the immediate vicinity of
$T_{c}$, where $\theta < \theta^{\ast} \sim 0.01$, the order of
the transition, by Eq. \eqref{vien9a}, is $1 / c_{1} \theta^{\ast}
\sim 100$, i.e. giant.  The order is non-universal, it depends on
the DNA parameters $T_{1}$, $T_{2}$, $\bar{x}$. The values of
$T_{1}$, $T_{2}$ depend on the ligands and their concentrations in
the DNA solutions \cite{Wartell, Azbel1973}, which may be
manipulated experimentally.  Non-universality in Eqs.
(\ref{vien9a}, \ref{vien9b}) is related to the competition of the
refractory and loop entropy terms in Eq. \eqref{vien5}, which
renormalizes the loop entropy, and thus the singularity. The width
of the transition is very small, yet macroscopic. The crossover
from Eq. \eqref{vien9b} to Eq. \eqref{vien9a} occurs when $( T_{c}
- T ) / T_{c} \sim 10^{-4}$.  Then $T_{c} - T \sim 0.01 K$ (cf
$\delta T_{c} =T_{c}  - \bar{T} \sim 3 K)$. In the approximation
of Eq. \eqref{vien6}, the probability density $P_{L}$ of a given
$L$ is $P_{L} =  M^{-1} (\ln L)^{1 / 2} L^{-1 - \delta} \exp (-
\phi L)$. So, by Eqs. (\ref{vien6}, \ref{vien9b}), $ \bar{L}
\propto 1 / \phi \propto \exp [ 1 / (T_{c} - T) ]$ exponentially
increases to $\bar{L} \sim 10^{40}$ at the crossover. Thus, even
in a solution with $\sim 10^{22}$ DNA nucleotide base pairs, all
DNA molecules completely unbind in the interval \eqref{vien9b}.
So, at a small, yet macroscopic distance $ \sim 0.01 K$ from
$T_{c}$, the effective long range interaction exceeds any
macroscopic size of the system. The system can no more be divided
into weakly interacting subsystems, thus the Gibbs distribution is
invalid, and thermodynamics must be calculated from the first
principles. The fraction of bounded sites is correspondingly small
there, and the observably quantity is the temperature of complete
melting of a finite DNA. If its length is $N$, then $\bar{L} = N$
at the temperature $T_{N}$, where
\begin{equation} \label{vien10}
   \theta_{N} = (T_{c} - T_{N}) / T_{c} \sim 1 / \ln N.
\end{equation}
The mean fluctuation $\Delta \theta_{N}$ of $\theta_{N}$ may be estimated from
$\bar{L} (\theta_{N} + \Delta \theta_{N}) - \bar{L} (\theta_{N}) = \Delta^{\ast} L (\theta_{N})$.
Similar to Eq. \eqref{vien4b}, $\Delta^{\ast} L \sim \bar{L}$  and thus \footnote{If DNA is not closed, then an
unbounded segment at a free DNA end does not yield the loop entropy.  This should be accounted for at the final
stages of unbinding of an open-ended (rather than closed as assumed in the Poland-Scheraga model) DNA.}
\begin{equation} \label{vien11}
   \Delta \theta_{N} / \theta_{N} \sim 1 / \ln N.
\end{equation}
Such fluctuation is macroscopic and easily observable. In DNA this situation is related to mesoscopic size
of base pairs (which yields large $J / k_{B} T \sim s \sim 10$, thus small $\theta^{\ast}$), and to DNA heterogeneity.
By Eq. \eqref{vien6}, heterogeneity effectively replaces fixed $c_{1}$ with $\delta$.
The latter decreases to $\delta_{c} \ll 1$ (at $T = T_{c}$) and scales the transition order with
$1 / \delta_{c}$ in Eq. \eqref{vien9a} and with $1 / \delta$ in Eq. \eqref{vien9b}.
This may be characteristic of any sufficiently strong long range interaction.

By Eq. \eqref{vien10}, natural DNA always yields $\theta \gg \theta^{\ast}$, i.e. the essential singularity \eqref{vien9b}
in $\theta \propto T_{c} - T$.  (This was predicted in ref. \cite{Azbel1973}).  By Eq. \eqref{vien8}, it proceeds
in the interval $T_{c} - T \sim 0.01 T_{c} \sim 3 K$. Sufficiently close to $\bar{T}$, the length $\bar{L}$
may reach the correlation length of the sequence. Then the distribution $w (l, x)$  becomes non-Gaussian.
This alters Eq. \eqref{vien5} and the melting curve $\phi (T)$.

Below $\bar{T}$ DNA is mostly bounded, and only anomalously fusible segments melt.
Their probability yields the equation which replaces Eq. \eqref{vien5}. Their melting proceeds in an entire
interval $\Delta T$. Until sufficiently high temperatures, when the number of segments, which melt nearly simultaneously,
becomes large, the DNA melting curve exhibits their successive melting. It is explicitly seen in experiments
\cite{Wartell, Azbel1973}. Thus, in a general case there are three distinctly different temperature intervals:
$\theta^{\ast} \sim 0.01$, i.e. $T_{c} - T \sim 0.03 K$;  $\theta \sim 1$, i.e. $T_{c} - T \sim 3 K$;
and $\Delta T \sim 40 K$.

A giant order transition \eqref{vien9a} may be observed only when the total number $N$ of base  pairs is much larger
than $\bar{L}$ at the crossover to Eq. \eqref{vien9b}. This implies $\ln N > 1 / \theta^{\ast}$.
Since, by  Eq. \eqref{vien8},  $\theta^{\ast} \sim 0.005 D^{-4 / 3}$,  so $D$ must be $< 0.03(\ln N)^{3 / 4}$.
On the other hand, the derivation of Eq. \eqref{vien6} implied the large renormalized term.
At the crossover this means $D > 0.03$.  In the interval $0.03 < D < 0.03(\ln N)^{3 / 4}$ non-universality of the
giant critical index in Eq. \eqref{vien9a} may be studied (e.g., via its dependence on $\Delta T$,
which changes together with the concentration of solvents in DNA solution \cite{Wartell}).

Presented theory may be numerically tested. Once the ground state is accurately determined analytically \cite{Azbel1973},
computer simulations allow for the study of its fluctuations.

The approach is applicable to other problems also.

To summarize. DNA unbinding with temperature proceeds from
piecewise melting of fusible domains, to essential singularity, to
giant ($\sim 1 / \theta^{\ast} > 100$) order phase transition. The
latter may be observed when the AT or GC concentration is between
$0.03$ and $0.03(\ln N)^{3 / 4}$, where $N$ is the total number of
nucleotide pairs. In the vicinity of complete melting the Gibbs
distribution is invalid, and thermodynamics must be calculated
from the first principles.

\begin{acknowledgments}
  Financial support from A. von Humboldt award and the J. and R. Meyerhoff chair is appreciated.
\end{acknowledgments}



\end{document}